%% file: IRS_pilot.tex
\documentclass[conference, letterpaper]{IEEEtran}
 \usepackage{amsmath}
 \usepackage{subfigure}
 \usepackage{graphicx,graphics,color,psfrag}
 \usepackage{cite,balance}
 \usepackage{amsmath,amssymb,amsfonts}
 \usepackage{caption}
 \usepackage{amsmath}
 \usepackage{amsthm}
\usepackage{textcomp}
 \allowdisplaybreaks
 \usepackage{algorithm}
  \usepackage{algorithmic}
 \usepackage{accents}
 \usepackage{amsthm}
 \usepackage{bm}
  \usepackage{url}
 \usepackage[english]{babel}
 \usepackage{multirow}
 \usepackage{enumerate}
 \usepackage{cases}
 \usepackage{stfloats}
 \usepackage{dsfont}
 \usepackage{color,soul}
 \usepackage{amsfonts}
 \usepackage{cite,graphicx}
 \usepackage{subfigure}
 \usepackage{fancyhdr}
 \usepackage{hhline}
 \usepackage{graphicx,graphics}
 \usepackage{array,color}
  \usepackage{bm}

\include{header}

\usepackage[top=0.73in, bottom=0.68in, left=0.585in, right=0.585in]{geometry}
\begin{document}
\captionsetup[figure]{name={Fig.}}
%\title{\huge 
%%\vspace{-10pt}
%%\title{\LARGE 
%UAV-Enabled Data Harvesting in Probabilistic LoS Channels}
%\author{Changsheng You and Rui Zhang   \thanks{\noindent C. You and R. Zhang are  with the Dept. of Electrical and Computer Engineering, National University of Singapore, Singapore (Email: eleyouc@nus.edu.sg, elezhang@nus.edu.sg).
% }}

\title{\huge 
Intelligent Reflecting Surface with Discrete Phase Shifts: Channel Estimation and Passive Beamforming} 
\author{\IEEEauthorblockN{Changsheng You, Beixiong Zheng, and Rui Zhang}
\IEEEauthorblockA{Department of Electrical and Computer Engineering,
National University of Singapore, Singapore\\
Email: \{eleyouc, elezbe, elezhang\}@nus.edu.sg}}
%\vspace{-10pt}
\maketitle

%\maketitle
%\vspace{-20pt}
\begin{abstract}
In this paper, we consider an intelligent reflecting surface (IRS)-aided single-user system where an IRS with  \emph{discrete} phase shifts  is deployed to assist the uplink communication. A practical transmission protocol is proposed to execute channel estimation
and passive beamforming successively.
% from a single-antenna user to a single-antenna access point (AP).  We assume that the channel state information (CSI) is \emph{unknown} at the AP and only limited channel training time is provided prior to data transmission. 
To minimize the mean square  error  (MSE) of channel estimation, we first formulate an optimization problem for designing the IRS reflection pattern in the training phase  under the constraints of unit-modulus, discrete phase, and full rank. This problem, however, is NP-hard and thus difficult to solve in general. As such, we propose a low-complexity yet efficient method to solve it sub-optimally, 
%called discrete Fourier transform \emph{(DFT)-Hadamard-based} reflection pattern design,
by constructing a \emph{near-orthogonal} reflection pattern based on either discrete Fourier transform \emph{(DFT)-matrix quantization} or \emph{Hadamard-matrix truncation}. 
%Note that the channel estimation error acts like interference in signal detection which varies with the IRS passive beamforming. 
% The MSE by using this method is observed to be closely determined by the orthogonality condition of the devised reflection pattern. 
Based on the estimated channel, we then formulate an optimization problem to maximize the achievable rate by designing the discrete-phase passive beamforming at the IRS with the training overhead and channel estimation error taken into account. To reduce the computational complexity of exhaustive search, we further propose a low-complexity successive refinement algorithm with a properly-designed initialization  to obtain a high-quality suboptimal solution.
% \emph{successive refinement} algorithm, which firstly determines the initial communication reflection coefficients using continuous relaxing followed by discrete quantization, and then iteratively optimizes the reflection coefficient of each sub-surface with those of the others being fixed. 
 Numerical results are presented to show the significant rate improvement of our proposed IRS training reflection pattern and passive beamforming designs as compared to other benchmark schemes.

\end{abstract}

%\begin{IEEEkeywords}
%Intelligent reflecting surface, limited training, discrete phase shift, channel estimation, reflection optimization.
%\end{IEEEkeywords}

\section{Introduction}
%Leveraging the recent advances in radio frequency (RF) micro electromechanical systems (MEMS) and metamaterial,
 \emph{Intelligent reflecting surface} (IRS), which is composed of  a vast number of passive reflecting elements, is capable of dynamically  controlling both the amplitude and phase of reflected signal and thus enables \emph{smart reconfiguration} of the radio environment\cite{qingqing2019towards,basar2019wireless}. Besides, IRS is \emph{cost-effective} in the sense that it does not require any active radio frequency (RF) chains for signal transmission/reception but simply relies on passive signal reflection, thus significantly reducing the cost and  energy consumption as compared to traditional active transceivers/relays. Moreover, IRS	can be easily attached to or removed from different objects (e.g., walls and ceilings), hence featuring extraordinary flexibility and compatibility for the practical deployment. These appealing advantages of IRS have spurred rapidly growing interests recently in designing   new IRS-aided wireless communication  systems by revamping classic  communication techniques such as multiple-input multiple-output (MIMO)\cite{zhang2019capacity}, orthogonal frequency division multiplexing (OFDM) \cite{yang2019intelligent}, and non-orthogonal multiple access (NOMA) \cite{yang2019intelligentnoma}, to name a few.

Most of the existing works on IRS (e.g., \cite{wu2019intelligent,huang2019reconfigurable}) have  assumed perfect channel state information (CSI) at the access point (AP) and focused on the joint design of active transmit beamforming at the AP and passive reflect beamforming at the IRS to maximize the system spectral/energy efficiency.
%to constructively align the IRS-reflected and non-IRS-reflected signals in phase at the user for rate enhancement. 
The acquisition of perfect CSI, however, is practically difficult to realize
in IRS-aided systems due to its large number of reflecting elements and lack of signal transmission/processing capabilities. 
To address this issue, the authors in \cite{zheng2019intelligent} proposed to \emph{group} IRS elements for reducing the channel estimation and passive beamforming complexity. Based on IRS-elements grouping, a  discrete Fourier transform (DFT)-based IRS reflection pattern was proposed to minimize the channel estimation error.
This work, however, only considered the continuous phase shifts for the ease of design, which are practically difficult to achieve for an IRS with typically a large number of reflecting elements, due to   the high cost for manufacturing each reflecting element with an infinite-level high-resolution phase shifter. Thus, it is more practical to consider  \emph{discrete} phase shifts at each reflecting element with a small number of controlling bits, e.g.,  $1$-bit for two-level ($0$ or $\pi$) phase shifts \cite{wu2019beamforming}. As such, beyond the existing works on IRS channel estimation assuming continuous phase shifts \cite{zheng2019intelligent,mishra2019channel,he2019cascaded}, two new questions arise when considering discrete phase shifts in the designs of channel estimation and passive beamforming in IRS-aided systems. Firstly,  it is unknown whether the DFT-based reflection pattern for the channel training that works for continuous phase shifts is still applicable to the case with only discrete phase shifts. Secondly, the effects of discrete phase shifts on the channel estimation error and passive beamforming performance are yet to be characterized.  
%\begin{itemize}
%\item[1)] 
%\item[2)] It is unknown what are the effects of the discrete phase shifts on the channel estimation error and passive beamforming.
%\end{itemize}

To address the above questions, we consider  in this paper an IRS-aided single-user system with discrete-phase reflecting elements that are grouped into a relatively small number of sub-surfaces.  A practical transmission protocol is proposed to execute channel estimation
and passive beamforming successively.
 To minimize the mean-square error (MSE) of channel estimation with limited training time, we formulate an optimization problem for designing the IRS reflection pattern in the training phase  under the constraints of unit-modulus, discrete phase, and full rank. This problem, however,  is NP-hard, whose optimal solution can only be obtained by the exhaustive search. To reduce the computational complexity, we first show that the simple DFT/Hadamard-based reflection pattern is  optimal in some special cases. Then, for other general cases, we propose a low-complexity yet efficient method to sub-optimally solve this problem, called \emph{DFT-Hadamard-based} reflection pattern design, which systematically constructs a \emph{near-orthogonal} reflection pattern based on  either \emph{DFT-matrix quantization} or \emph{Hadamard-matrix truncation}. Due to the discrete-phase constraint, the designed  reflection pattern cannot always preserve matrix orthogonality as in the case of continuous phase shifts, thus leading to the enhancement  and correlation of the channel estimation error.
 Particularly, the correlated channel estimation error incurs complicated  interference in data transmission,  which varies with the IRS passive beamforming. We thus formulate an optimization problem to maximize the  achievable rate in data transmission  by designing the passive beamforming with the training overhead and the correlated channel estimation error taken into account. 
To reduce the computational complexity for the optimal solution based on exhaustive search, we propose a low-complexity \emph{successive refinement} algorithm with a properly-designed initialization to obtain a high-quality suboptimal solution.
Numerical results verify the effectiveness of our proposed IRS training reflection pattern and passive beamforming designs.

%\emph{Notations:} The superscripts $(\cdot)^{T}$, $(\cdot)^{H}$, $(\cdot)^{\dag}$, and $(\cdot)^{-1}$ denote respectively the operations of transpose, Hermitian transpose, conjugate, and matrix inversion. $\diag(\cdot)$ denotes a square diagonal matrix with the elements on its main diagonal, $[\cdot]_{i,j}$ is the ($i,j$)-th element of a matrix, $\angle(\cdot)$ denotes the phase of a complex number, and $\rank(\cdot)$ and $\tr(\cdot)$ represent the matrix rank and trace, respectively.
%\vspace{-5pt}
\section{System Model}\label{Sec:Model}

\begin{figure}[t]
\vspace{2pt}
\begin{center}
\includegraphics[height=4.4cm]{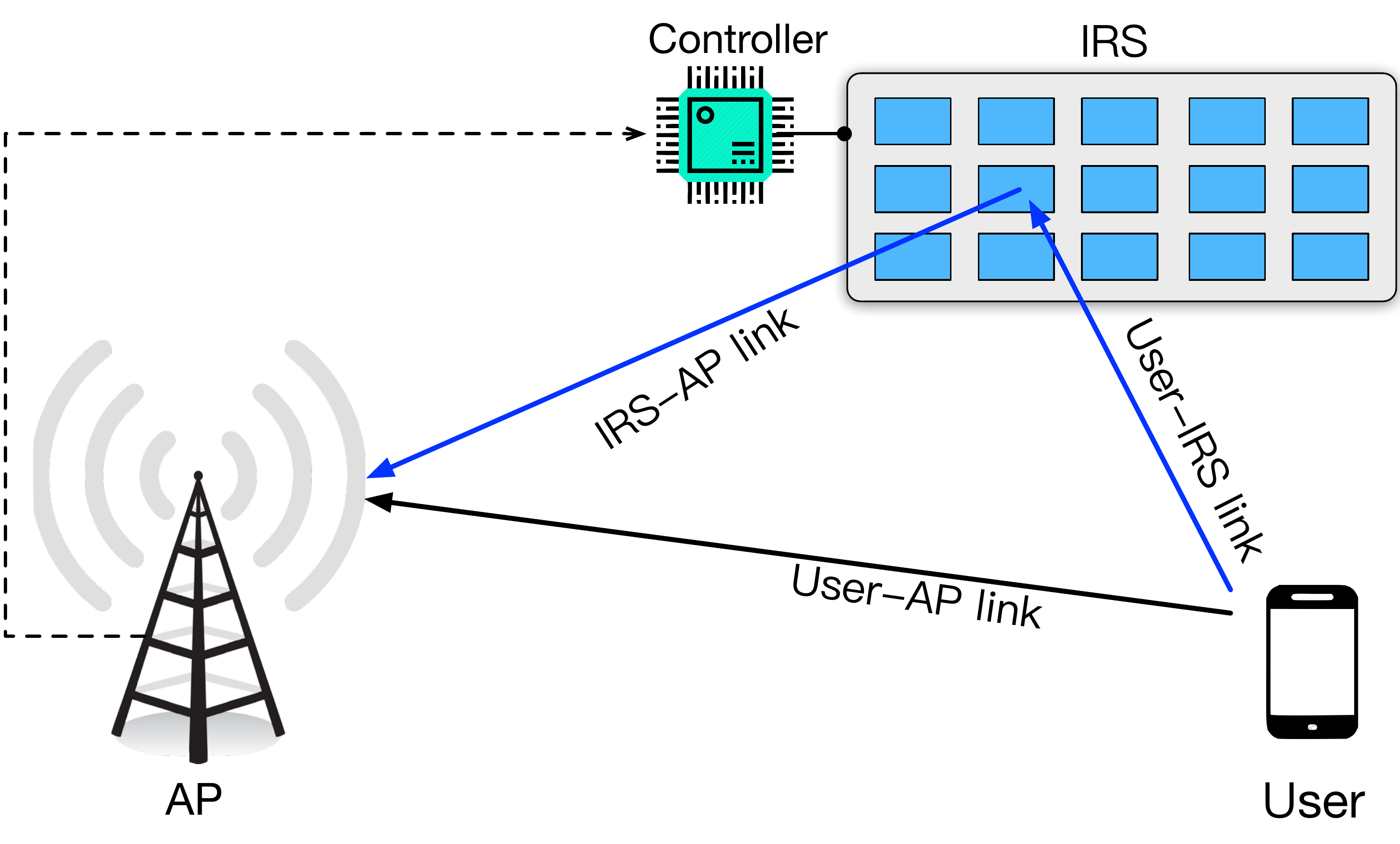}
\caption{An IRS-aided single-user uplink communication system.}
\label{Fig:Syst}
\end{center}
\end{figure}

Consider an IRS-aided single-user communication system as illustrated in Fig.~\ref{Fig:Syst}, where an IRS composed of a large number of passive reflecting elements is deployed in proximity  to a user for assisting its uplink data transmission to an AP, both of which are equipped with a single antenna. The IRS is attached with a smart controller, responsible for dynamic adjustment of the amplitude and/or phase shift at each reflecting element as well as the information exchange between the IRS and AP via a separate reliable wireless link.

\underline{Channel model:} For exploiting the high channel correlation between adjacent reflecting elements, we recall the IRS-elements grouping scheme in \cite{zheng2019intelligent} to reduce the channel estimation overhead, where $N$ reflecting elements at the IRS  are divided into $M$ groups, each called a sub-surface consisting of $N/M$ adjacent elements  sharing a \emph{common} reflection coefficient. 
For a typically low-mobility user  served by the IRS, we assume quasi-static block fading channels and focus on the uplink communication in one particular fading block, where the associated channels remain constant within this fading block. 
Let $h_{\rm{UA}}\in\mathbb{C}$, $\boldsymbol{h}_{\rm UI}\in\mathbb{C}^{M\times1}$, and $\boldsymbol{h}_{\rm{IA}}^{H}\in\mathbb{C}^{1\times M}$ denote respectively the baseband equivalent channels of the user-AP link, and the  per-sub-surface based  user-IRS and IRS-AP links. 
%associated with the sub-surfaces.
%, where  $N$ is the number of IRS reflecting elements. 
Based on \cite{zheng2019intelligent}, the IRS reflection coefficients of its sub-surfaces can be represented by a diagonal matrix, denoted by $\boldsymbol{\Omega}=\diag(\beta_1 e^{j\omega_1}, \cdots, \beta_m e^{j\omega_m}, \cdots, \beta_M e^{j\omega_M})$, where $\beta_m\in[0,1]$ and $\omega_m\in[0, 2\pi)$ respectively denote the common reflection amplitude and phase shift of the elements in the $m$-th sub-surface. In practice, the phase shift of each reflecting element can only take a finite number of discrete values due to hardware limitations \cite{qingqing2019towards,wu2019beamforming}. Let  $b$ denote the number of bits used to uniformly quantize the continuous phase shift  in $[0, 2\pi)$. Then the set of discrete phase shifts of each sub-surface can be represented by  $\mathcal{F}=\{0, \Delta\omega, \cdots, (K-1)\Delta\omega\}$,
where $\Delta\omega=2\pi/K$ and $K=2^{b}$ is the number of discrete phase-shift levels. To ease  the design of reflection coefficients and enhance the reflected signal power, we consider full-on reflection at the IRS, where the reflection amplitude of each sub-surface is set to be the maximum, i.e., $\beta_m=1, m=1, \cdots, M$. 
Moreover, we assume that the signals reflected by the IRS more than once are of negligible power due to the high path loss and thus can be ignored. Similar to \cite{wu2019intelligent}, the equivalent channel from the user to AP is given  by
\vspace{-5pt}
\begin{equation}
g= \boldsymbol{h}_{\rm{IA}}^{H} \boldsymbol{\Omega} \boldsymbol{h}_{\rm{UI}}+ h_{\rm{UA}}.\label{Eq:Chan}
\end{equation}
Let $\boldsymbol{h}_{\rm{R}}\triangleq\diag(\boldsymbol{h}_{\rm{IA}}^{H}) \boldsymbol{h}_{\rm{UI}}$ denote the cascaded user-IRS-AP channel in the absence of phase shifts, and $\boldsymbol{{\bar\theta}}^{H}\triangleq \l[e^{j\omega_1}, \cdots, e^{j\omega_M}\r]$ denote the sub-surface  reflection vector. Then the equivalent channel given in \eqref{Eq:Chan} can be rewritten as $g= \boldsymbol{{\bar\theta}}^{H}\boldsymbol{h}_{\rm{R}}+ h_{\rm{UA}}$, which can be further simplified as
\vspace{-5pt}
\begin{align}
g=\boldsymbol{\theta}^{H} \boldsymbol{h},\label{Eq:EffecChan}
\end{align}
where $\boldsymbol{\theta}^{H}\triangleq [1, \boldsymbol{{\bar\theta}}^{H}]$ and  $\boldsymbol{h}^{H}\triangleq[h_{\rm{UA}}^{\dag}, \boldsymbol{{h}}_{\rm{R}}^{H}]$
denote respectively the (extended) IRS reflection coefficients  and channels accounting for  both the direct and cascaded user-IRS-AP links.

%\begin{figure}[t]
%%\vspace{2pt}
%\begin{center}
%\includegraphics[height=1.8cm]{./FigProtocal.pdf}
%\caption{A two-phase protocol for the IRS-aided uplink communication.}
%\label{Fig:Proto}
%\end{center}
%\end{figure}

%\subsection{Transmission Protocol}
\underline{Transmission protocol:}
To achieve passive beamforming gain provided by the IRS, existing works assume perfect CSI at the AP/IRS, which is practically difficult to achieve.  
%As the perfect CSI in the IRS-aided system is practically difficult to obtain due to the massive number of reflecting elements,
 Therefore, we consider a practical transmission protocol for successive channel training and data transmission. Specifically, 
% As illustrated in Fig.~\ref{Fig:Proto},
  a transmission frame of $T_0$ symbol durations is partitioned into two sub-frames, corresponding to the first sub-frame of $(M+1)$ symbol durations for channel estimation and the second one of the remaining $T_0-(M+1)$ symbol durations for data transmission. Note that the $(M+1)$ training symbols are used to estimate the channels of the direct link and the cascaded user-IRS-AP  links associated with the $M$ sub-surfaces.

During the channel training, the user consecutively sends $(M+1)$ pilot symbols to the AP, where  the  IRS  reflection coefficients are properly set to assist the channel estimation at the AP. For each symbol duration $m\in\mathcal{M}\triangleq\{1, \cdots, M+1\}$, let  $x_p[m]\in \mathbb{C}$ and $\boldsymbol{\theta}_p[m]$ denote the  transmitted pilot signal  and extended IRS reflection coefficients. Then the baseband received signal, $y_p[m]$, is given by
\vspace{-5pt}
\begin{equation}
y_p[m] = x_p[m] g_p[m]  +z_p[m], ~~m \in \mathcal{M},\label{Eq:TrainRec}
\end{equation}
where $g_p[m]=\boldsymbol{\theta}_p^{H}[m] \boldsymbol{h}$ and $z_p[m]$ is the additive white Gaussian noise  at the receiver with zero mean and variance $\sigma^2$. By stacking $(M+1)$ consecutive received signals, i.e. $\boldsymbol{y}_p=[y_p[1], \cdots, y_p[M+1]]^{T}$, the received signal vector during the channel training can be compactly written as
\vspace{-5pt}
\begin{equation}
\boldsymbol{y}_p= \boldsymbol{X}_p \boldsymbol{g}_p  +\boldsymbol{z}_p,
\label{Eq:TrainRecComp}
\end{equation}
where $\boldsymbol{X}_p\triangleq\diag(x_p[1], \cdots, x_p[M+1])$, $\boldsymbol{g}_p\triangleq[g_p[1],\cdots, g_p[M+1]]^{T}$, and $\boldsymbol{z}_p\triangleq[z_p[1], \cdots, z_p[M+1]]^{T}$.

Upon receiving $\boldsymbol{y}_p$, the AP estimates the extended channel $\boldsymbol{h}$ (with the details given later in Section~\ref{Sec:ChanEst}),  based on which, it optimizes the passive beamforming with discrete phase shifts at the IRS, i.e.  $\boldsymbol{\theta}^{H}$, to maximize the achievable rate for data transmission (see Section~\ref{Sec:RefOpt}). The optimized passive beamforming coefficients  are then fed back to the IRS controller, based on which the IRS controller tunes the phase shift at each reflecting element accordingly to assist the data transmission. 
%The detailed designs for the two phases are elaborated in the following sections.
\section{Channel Estimation}\label{Sec:ChanEst}
In this section, an optimization problem is formulated and solved  to minimize the MSE of the least-square (LS) channel estimation by designing the IRS reflection pattern.  
%To efficiently solve this NP-hard problem, we propose a systematic method to obtain a high-quality sub-optimal solution to it with low complexity.
\subsection{Problem Formulation}
First,  an (extended) IRS reflection pattern $\boldsymbol{\Theta}_p$ is defined as a matrix that stacks the $(M+1)$ (extended) reflection vectors, i.e.,
 $\boldsymbol{\Theta}_p \triangleq [\boldsymbol{\theta}_p[1], \cdots, \boldsymbol{\theta}_p[M+1]]^H$.
   Then $\boldsymbol{g}_p=\boldsymbol{\Theta}_p\boldsymbol{h}$ and the  LS estimation of $\boldsymbol{g}_p$, denoted by $\boldsymbol{{\tilde g}}_p$, can be derived from \eqref{Eq:TrainRecComp} as follows:
\begin{align}
\boldsymbol{{\tilde g}}_p &= \boldsymbol{X}_p^{-1}\boldsymbol{y}_p= \boldsymbol{g}_p +\boldsymbol{X}_p^{-1}\boldsymbol{z}_p.
\end{align}
Thus, if  $\boldsymbol{\Theta}_p$ is of full-rank, the estimation of $\boldsymbol{ h}$ is given by
\begin{equation}
\boldsymbol{{\tilde h}}
%\begin{bmatrix}
%   {\tilde h}_{\rm{UA}} \\
%   \boldsymbol{{\tilde{ h}}}_{\rm{R}}
%  \end{bmatrix}
  =\boldsymbol{\Theta}_p^{-1}\boldsymbol{{\tilde g}}=\boldsymbol{{ h}}+\boldsymbol{{ h}}_e,\label{Eq:ChanEsti}
\end{equation}
where $\boldsymbol{{ h}}_e\triangleq\boldsymbol{\Theta}_p^{-1}\boldsymbol{X}_p^{-1}\boldsymbol{z}_p$  denotes the channel estimation error for $\boldsymbol{{ h}}$. As such, the MSE of the above  LS channel estimation is given by
\begin{align}
{\rm MSE}&=\mathbb{E}\l[|| \boldsymbol{{h}}-\boldsymbol{{\tilde h}}||
  ^2\r]=\mathbb{E}\l[||\boldsymbol{{ h}}_e||
  ^2\r]\nn\\
  &=\mathbb{E}\l[\tr\l( \boldsymbol{\Theta}_p^{-1}\boldsymbol{X}_p^{-1}\boldsymbol{z}_p \boldsymbol{z}_p^{H}(\boldsymbol{X}_p^{-1})^{H} (\boldsymbol{\Theta}^{-1}_p)^{H}\r)\r]\nn\\
  &=\frac{\sigma^2}{P_t}\tr((\boldsymbol{\Theta}^{H}_p \boldsymbol{\Theta}_p)^{-1}),
\end{align} 
where $P_t$ denotes the transmit power of the user.

Accounting for the IRS discrete phase shifts and the feasibility of the LS estimation method,  the constraints for the feasible IRS reflection pattern, $\boldsymbol{\Theta}_p$, are listed as follows. 
\begin{itemize} 
\item[1)] The first column of  $\boldsymbol{\Theta}_p$ should be an all-one vector due to the estimation for the direct link without phase shift, i.e., 
\begin{equation}
[\Theta_p]_{i,1}=1,\quad 1\le i\le M+1.\label{Eq:RefCons1}
\end{equation} 
\item[2)] The entries of the IRS reflection pattern should satisfy the constraints of unit-modulus and discrete phase, i.e., 
\begin{align}
|[\Theta_p]_{i,j}|=1, ~~~~~ 1\le i\le M+1, 2\le j\le M+1, \label{Eq:RefCons2}\\
\angle[\Theta_p]_{i,j}\in\mathcal{F}, ~~~1\le i\le M+1, 2\le j\le M+1. \label{Eq:RefCons3}
\end{align}
\item[3)] The IRS reflection pattern should be of full-rank, i.e., 
\begin{equation}
\rank(\boldsymbol{\Theta}_p)=M+1.\label{Eq:RefCons4}
\end{equation}
\end{itemize} 
Under the above constraints, the optimization problem for minimizing the MSE of channel estimation is formulated as
\vspace{-5pt}
\begin{subequations}
\begin{align}
({\bf P1}):~~\min_{\boldsymbol{\Theta}_p} ~~ &\frac{\sigma^2}{P_t}\tr\l((\boldsymbol{\Theta}^{H}_p \boldsymbol{\Theta}_p)^{-1}\r) \nn \\  
~~~~\text{s.t.}~~~
& \eqref{Eq:RefCons1}-\eqref{Eq:RefCons4}.\nn
\end{align}
\end{subequations}
\subsection{Proposed Reflection Pattern Design}
First, it can be easily verified that problem (P1) is always feasible, since there exists a \emph{naive} reflection pattern that satisfies all the constraints in \eqref{Eq:RefCons1}-\eqref{Eq:RefCons4},  regardless of the phase-shifter resolution and the channel training time. We denote it by $\boldsymbol{{\bar\Theta}}_p$, whose entries are given by
\vspace{-5pt}
% $\boldsymbol{{\bar\Theta}}_p$ is given by
\begin{align}\label{Eq:NaiveSch}
[\boldsymbol{{\bar\Theta}}_p]_{i,j}=\begin{cases}
-1,&~i=j~\text{and}~i\neq 1,\\
1,&~\text{otherwise}.\\
\end{cases}
\end{align}

However, despite its feasibility, the objective function of (P1) is non-convex due to the inverse operation as well as  the non-convex  constraints of full rank and unit-modulus. In addition, the phase shifts of the IRS reflection pattern are restricted in a finite number of discrete values, rendering problem (P1) an NP-hard problem to solve. Numerically, the optimal solution to problem (P1) can be obtained by an exhaustive search over all possible reflection patterns that satisfy the constraints in \eqref{Eq:RefCons1}-\eqref{Eq:RefCons3}, with the complexity of order  $\mathcal{O}(2^{bM\times(M+1)})$, and then selecting the one with full rank and achieving the minimum MSE (MMSE). Note that the optimal solution may \emph{not be unique}. The computational complexity of the exhaustive search, however, is practically prohibitive, since it increases exponentially with $M$ and/or $b$.
%when the channel training time (i.e., $M+1$) is relatively long and/or the discrete phase shifters are of high resolution (i.e., $b$ is large).

To address this issue, we first obtain the optimal reflection pattern for (P1) in some special cases with respect to (w.r.t.) $b$ and  $M$.  Then for other general cases, we propose a low-complexity algorithm to obtain a high-quality suboptimal  solution to problem (P1).
To this end, we first introduce two structured matrices: the DFT matrix and the Hadamard matrix.  Specifically, an $(M+1)\times (M+1)$ DFT matrix, denoted by $\bar{\boldsymbol{D}}_{M+1}$, is an orthogonal matrix whose entries are given by
\vspace{-5pt}
\begin{equation}
\l[\bar{\boldsymbol{D}}_{M+1}\r]_{i,j}=e^{-j\frac{2\pi (i-1) (j-1)}{M+1}}, ~~~1\le i, j\le M+1.
\end{equation}
On the other hand, a Hadamard matrix is also an orthogonal matrix but its entries are either $+1$ or $-1$.  For example, a $4\times4$ Hadamard matrix, denoted by $\bar{\boldsymbol{H}}_{\rm 4}$, is given by
\vspace{-5pt}
\begin{equation}
\bar{\boldsymbol{H}}_4=\begin{bmatrix}
   1  & 1 & 1& 1 \\
   1  & -1 & 1& -1 \\
   1  & 1 & -1& -1 \\
   1 & -1 & -1& 1
  \end{bmatrix}.\label{Eq:exaInv}
\end{equation}
Note that an $(M\!+\!1)\!\times \!(M\!+\!1)$ Hadamard matrix exists only if $$M+1\in \mathcal{U}\triangleq\{u| u=2 ~\text{or}~u=4r, r\in\mathbb{Z}^{+}\}.$$
% denoted by $\bar{W}_{\rm D}$,  is an orthogonal matrix whose entries are either $+1$ and $-1$ and whose row are mutually orthogonal. 
%From the above definitions, we can easily verify that both the DFT and Hadamard matrices satisfy the constraints in \eqref{Eq:RefCons1}-\eqref{Eq:RefCons4}. 
The following proposition gives the optimal reflection pattern solution to (P1) in two special cases.
%By leveraging their \emph{row-orthogonality}, we can easily derive the following lemma.

%  The following lemma gives the optimal solution to problem (P3).
\begin{proposition}\label{Lem:ContCE}\emph{For IRS with equally-spaced discrete phase shifts, the optimal reflection pattern for problem (P1) in the following two cases are given as follows.
\begin{itemize}
\item[1)] If $M+1\in \{2^{c}| c=1, 2, \cdots, b\}$, the DFT matrix $\bar{\boldsymbol{D}}_{M+1}$ is an optimal reflection pattern.
%two optimal reflection patterns are given by 
%\begin{equation}
%\boldsymbol{\Theta}_p^*=\bar{\boldsymbol{D}}_{M+1}~~\text{and}~~\boldsymbol{\Theta}_p^*=\bar{\boldsymbol{H}}_{M+1}.
%\end{equation}
\item[2)] If $M+1\in \mathcal{U}$, the Hadamard matrix $\bar{\boldsymbol{H}}_{M+1}$ is an optimal reflection pattern.
%  is $\boldsymbol{\Theta}_p^*=\bar{\boldsymbol{H}}_{M+1}$.
\end{itemize}
}
\end{proposition}
\emph{Sketch of Proof:} First, it can be shown that if  there exists an orthogonal reflection pattern, i.e., $\boldsymbol{\Theta}_p^H \boldsymbol{\Theta}_p=(M+1)\boldsymbol{I}$, satisfying all the constraints in \eqref{Eq:RefCons1}-\eqref{Eq:RefCons4}, then it is an optimal solution to problem (P1). Second, we can obtain the conditions where the orthogonal DFT and Hadamard matrices  satisfy the above-mentioned constraints, leading to the desired results.
 \hfill $\Box$

%Based on the optimal reflection pattern for the IRS with continuous phase shift as given Proposition~\ref{Lem:ContCE}, 
For other cases, in general, it is unknown whether there exists an orthogonal matrix satisfying all the constraints in \eqref{Eq:RefCons1}-\eqref{Eq:RefCons4}, which makes it hard to characterize the structure of the optimal solution to (P1). Thus we propose a \emph{novel} low-complexity  method, called \emph{DFT-Hadamard-based}  reflection pattern design, to obtain a suboptimal  solution to problem (P1). Basically, our proposed design systematically constructs a \emph{near-orthogonal} reflection pattern by performing DFT-matrix quantization for $b\ge 2$, and Hadamard-matrix truncation for $b=1$.
 The rationalities and detailed construction are elaborated as follows.

\begin{itemize}
\item[1)] \textbf{DFT-based reflection pattern for $b\ge 2$}:  Our goal is to construct a quantized DFT matrix $\boldsymbol{D}_{M}$ for any $M$, such that it features \emph{near-orthogonality} in the sense that each entry has a value closest to that of the corresponding DFT matrix, but with the phase shift constrained in the feasible set $\mathcal{F}$.  Mathematically, we have  $\l[\boldsymbol{D}_{M+1}\r]_{i,j}=e^{j\phi_{i,j}}$, where $$\phi_{i,j}=\arg\min_{\phi_{i,j}\in\mathcal{F}}\l| e^{j\phi_{i,j}}- e^{-j\frac{2\pi (i-1) (j-1)}{M+1}}\r|.$$
Such a quantized DFT matrix, however, can no longer preserve matrix invertibility for the IRS with any resolution of phase shifters. By extensive simulations, we observe that the quantized DFT matrix is always invertible for $b\ge2$ and achieves an MSE close to that of the continuous phase shifts when $b$ is large. While, for the IRS with $b=1$, i.e., $1$-bit phase shifters, the proposed quantized-DFT reflection pattern, $\boldsymbol{D}_{M+1}$, is mostly \emph{irreversible} for different training time. For instance, we observe that  for $1\le M\le 50$,  $\boldsymbol{D}_{M+1}$ is invertible only when $M+1\in\{2, 4, 8, 16, 32\}$, for which each of the quantized DFT matrices reduces to a Hadamard matrix with the same dimension. Thus, we resort to a Hadamard-based scheme as described below for designing the reflection pattern when $b=1$.
% The design for the case with $1$-bit phase shiftscan be resorted to the next scheme.
\item[2)] \textbf{Hadamard-based reflection pattern for $b=1$}:  For the IRS with $1$-bit phase shifters,  by  leveraging the orthogonality of the Hadamard matrix, we propose to construct a \emph{truncated Hadamard} matrix for obtaining a \emph{near-orthogonal} reflection pattern $\boldsymbol{H}_{M+1}$ as follows. First, find an $\ell\times \ell$ 
legitimate Hadamard matrix $\bar{\boldsymbol{H}}_{\ell}$ that has the smallest dimension $\ell$ while satisfying $\ell\ge M+1$. Then, truncate $\bar{\boldsymbol{H}}_\ell$ by preserving only the entries in the first $M+1$ rows and first $M+1$ columns. Mathematically, we have $$[\boldsymbol{H}_{M+1}]_{i,j}=[\bar{\boldsymbol{H}}_\ell]_{i,j}, 1\le i, j\le M+1.$$ 
\end{itemize}
It is worth mentioning that  the optimal reflection pattern in the special cases given in Proposition~\ref{Lem:ContCE} are also special cases of  the above proposed DFT-Hadamard-based design. Moreover, note that  the MSE of the proposed scheme is dependent on the designed   reflection pattern  due to its \emph{non-orthogonality} in general, which is in sharp contrast to the case with continuous phase shifts for which the MMSE is a constant  given by \cite{zheng2019intelligent}
\begin{equation}
\text{MMSE} =\frac{\sigma^2}{P_t}\tr(((M+1)\boldsymbol{I})^{-1})=\frac{\sigma^2}{P_t}.
\end{equation}
%independent of the DFT-based reflection pattern . 

\begin{figure}[t]
\vspace{4pt}
\begin{center}
\includegraphics[height=4.5cm]{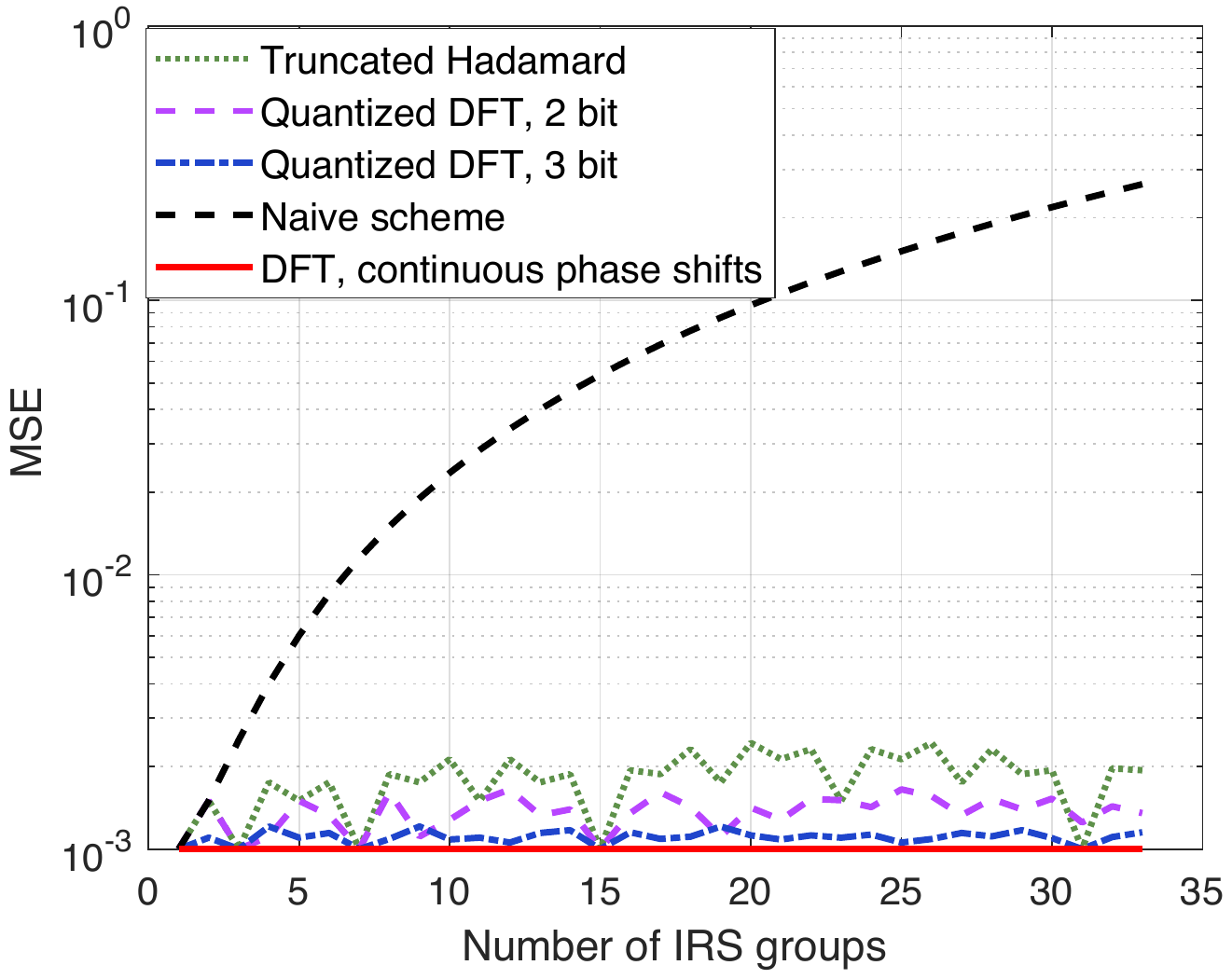}
\caption{MSE comparison between the proposed reflection pattern and the naive scheme for $1\le M\le 33$ with $P_t/\sigma^2=30$ dB.}
\label{FigMSE}
\end{center}
\end{figure}

Fig.~\ref{FigMSE} compares the MSE performance of the proposed DFT-Hadamard-based reflection pattern and the naive scheme whose entries are given in \eqref{Eq:NaiveSch}. 
First, it is observed that the MSE of the naive scheme monotonically increases with the number of IRS groups since the corresponding reflection pattern becomes more ill-conditioned. In contrast, our proposed scheme has much smaller MSE, albeit  fluctuating with the number of IRS groups. Specifically, its MSE touches the lower bound with  continuous phase shifts when $M+1\in\mathcal{U}$, since the corresponding reflection pattern reduces to a standard orthogonal Hadamard matrix. 
In addition, one interesting observation is that, although $\bar{\boldsymbol{D}}_{16}$ is not an optimal reflection pattern for the IRS with $2$-bit phase shifters since some of its phase shifts are out of the feasible phase-shift set (see Proposition~\ref{Lem:ContCE}), the quantized DFT-matrix is indeed a feasible and orthogonal matrix  and thus achieves the lower bound of the MSE.

\vspace{-5pt}
\section{Passive Beamforming Optimization}\label{Sec:RefOpt}
Based on the estimated channel,  we optimize the passive beamforming at the IRS in this section for maximizing the achievable rate for data transmission by taking into account  the \emph{reflection-pattern-dependent} channel estimation error. 
\vspace{-5pt}
\subsection{Problem Formulation}
In data transmission, the signal detection procedure at the AP is based on the estimated channel $\boldsymbol{{\tilde h}}$. The received signal at the AP is rewritten as
\vspace{-5pt}
% is the channels. Then the baseband received signal can be written by 
\begin{align}
y &= x \boldsymbol{\theta}^{H} \boldsymbol{h}  +z\nn\\
&=x \boldsymbol{\theta}^{H} (\boldsymbol{{\tilde h}}-\boldsymbol{{ h}}_e)  +z\nn\\
&=x \boldsymbol{\theta}^{H} \boldsymbol{{\tilde h}}-x \boldsymbol{\theta}^{H}\boldsymbol{{ h}}_e  +z,
\end{align}
where $ x \boldsymbol{\theta}^{H}\boldsymbol{{ h}}_e$ is the additional  interference due to the channel estimation error. Then the average achievable rate in bits/second/Hertz (bps/Hz) is given by \cite{samardzija2003pilot}
\vspace{-5pt}
\begin{align}
\!\!\!\!\!R=\frac{T_0-(M+1)}{T_0}\log_2\l(1+\frac{P_t|\boldsymbol{\theta}^{H}\boldsymbol{{\tilde h}}|^2}{\Gamma(P_t\mathbb{E}\l[|\boldsymbol{\theta}^{H}\boldsymbol{{ h}}_e|^2\r]+\sigma^2)}\r),\label{Eq:Rate}
\end{align} 
where $\Gamma\ge1$ stands for the achievable rate gap due to a practical modulation and coding scheme and $R$ is determined by 
%As such, maximizing the average achievable rate in \eqref{Eq:Rate} is equivalent to maximizing 
the following signal-to-interference-plus-noise ratio (SINR):
%where the signal-to-interference-plus-noise ratio (SINR) is
\begin{align}
\gamma(\boldsymbol{\theta})&=\frac{P_t|\boldsymbol{\theta}^{H}\boldsymbol{{\tilde h}}|^2}{P_t\mathbb{E}\l[|\boldsymbol{\theta}^{H}\boldsymbol{{ h}}_e|^2\r]+\sigma^2}.\label{Eq:SINR}
\end{align} 
By defining $\boldsymbol{{\tilde H}}\triangleq\boldsymbol{{\tilde h}}\boldsymbol{{\tilde h}}^{H}$ and
\vspace{-10pt} 
\begin{equation}
\boldsymbol{{R}}\triangleq\mathbb{E}[\boldsymbol{{ h}}_e\boldsymbol{{ h}}_e^{H}]=
%\boldsymbol{\Theta}_p^{-1}\boldsymbol{z}_p\boldsymbol{z}_p^H (\boldsymbol{\Theta}_p^{-1})^H=
\frac{\sigma^2}{P_t} \underbrace{(\boldsymbol{\Theta}_p^H \boldsymbol{\Theta}_p)^{-1}}_{\boldsymbol{{R}}_p},
\vspace{-15pt}
\end{equation}
 the SINR in \eqref{Eq:SINR} can be rewritten as
 \vspace{-5pt}
\begin{align}
\gamma(\boldsymbol{\theta})&=\frac{P_t\boldsymbol{\theta}^{H}\boldsymbol{{\tilde H}}\boldsymbol{\theta}}{\sigma^2(\boldsymbol{\theta}^{H}\boldsymbol{{R}}_p\boldsymbol{\theta}+1)}.\label{Eq:SINRNew}
\end{align} 
%Note that 
Note that $\boldsymbol{{R}}_p$ depends on the training reflection pattern of the IRS. From the above, the optimization problem for maximizing the average achievable rate in \eqref{Eq:Rate} under the constraints of unit-modulus and discrete phase  is equivalent to the optimization problem given below for the SINR maximization (by dropping the constant scaling factor $P_t/\sigma^2$).
\vspace{-5pt}
%Then the problem can be formulated as follows.
\begin{subequations}
\begin{align}
({\bf P2}):~\max_{\boldsymbol{\theta}} ~~&\frac{\boldsymbol{\theta}^{H}\boldsymbol{{\tilde H}}\boldsymbol{\theta}}{\boldsymbol{\theta}^{H}\boldsymbol{{R}}_p\boldsymbol{\theta}+1}& \nn \\  
\text{s.t.}~~
& |\theta_m|=1, ~\qquad\qquad m=1, \cdots, M+1,\label{Eq:P2Unit}\\
& \angle{\theta_1}=0,\angle{\theta_m}\in\mathcal{F}, m=2, \cdots, M+1.\label{Eq:P2Phase}
\end{align}
\end{subequations}
\subsection{Proposed Algorithms for Problem (P2)}
Problem (P2) is a non-convex optimization problem due to the constrains of unit-modulus and discrete phase. Since the discrete phase shifts are constrained in a finite set  $\mathcal{F}$,  the optimal solution can be obtained by  the exhaustive search, for which the complexity is of order $\mathcal{O}(2^{bM})$, which increases exponentially with $bM$.  
To reduce the computational complexity, we propose in this subsection an efficient \emph{successive refinement} algorithm with a properly-designed initialization to obtain a high-quality suboptimal solution to problem (P2).

First, we relax the constraint of discrete phase in \eqref{Eq:P2Phase} of problem (P2) and denote the resultant problem as problem (P3).  For this problem, 
%Observe that if $\boldsymbol{{\theta}}$ $\boldsymbol{{\bar\theta}}=\boldsymbol{{\theta}}e^{j\delta}$
 we define $\boldsymbol{\Phi}\triangleq\boldsymbol{\theta}\boldsymbol{\theta}^{H}$, which satisfies $\boldsymbol{\Phi}\succeq\boldsymbol{0}$ and $\rank(\boldsymbol{\Phi})=1$. Then we have  $\boldsymbol{\theta}^{H}\boldsymbol{{\tilde H}}\boldsymbol{\theta}=\tr(\boldsymbol{{\tilde H}}\boldsymbol{\theta}\boldsymbol{\theta}^{H})=\tr(\boldsymbol{{\tilde H}}\boldsymbol{\Phi})$ and $\boldsymbol{\theta}^{H}\boldsymbol{{R}}_p\boldsymbol{\theta}=\tr(\boldsymbol{{R}}_p\boldsymbol{\Phi})$.  By relaxing the non-convex rank-one constraint, problem (P3) is  transformed to
 \vspace{-5pt}
\begin{subequations}
\begin{align}
({\bf P4}):~~\max_{\boldsymbol{\Phi}} ~~&\frac{\tr(\boldsymbol{{\tilde H}}\boldsymbol{\Phi})}{\tr(\boldsymbol{{R}}_p\boldsymbol{\Phi})+1}& \label{Eq:P4Obj} \\  
~~~~\text{s.t.}~~
& \boldsymbol{\Phi}\succeq\boldsymbol{0},\\
& [\boldsymbol{\Phi}]_{m,m}=1, ~~~m=1, \cdots, M+1.\label{Eq:P4Diag}
\end{align}
\end{subequations}
Problem (P4) is still non-convex since the objective function is non-convex over $\boldsymbol{\Phi}$. To address this issue, we apply the Charnes-Cooper transformation to reformulate problem (P4) \cite{charnes1962programming}. To be specific, we define
\vspace{-5pt}
\begin{align}
\boldsymbol{\Psi}=\frac{\boldsymbol{\Phi}}{\tr(\boldsymbol{{R}}_p\boldsymbol{\Phi})+1},~~t=\frac{1}{\tr(\boldsymbol{{R}}_p\boldsymbol{\Phi})+1}.
\end{align}
As such, we have $\boldsymbol{\Phi}=\frac{\boldsymbol{\Psi}}{t}$ and  $\tr(\boldsymbol{{R}}_p\boldsymbol{\Psi})+t=1$.
%, and $\boldsymbol{\Phi}$ can be represented by $\boldsymbol{\Phi}=\frac{\boldsymbol{\Psi}}{t}$ and 
Consequently, problem (P4) is equivalent to the following problem.
\vspace{-5pt}
\begin{subequations}
\begin{align}
({\bf P5}):~~\max_{\boldsymbol{\Psi},t} ~~&\tr(\boldsymbol{{\tilde H}}\boldsymbol{\Psi})& \nn \\  
~~~~\text{s.t.}~~~
& \tr(\boldsymbol{{R}}_p\boldsymbol{\Psi})+t=1,\\
&\boldsymbol{\Psi}\succeq\boldsymbol{0},\nn\\
& [\boldsymbol{\Psi}]_{m,m}=t, ~~~m=1, \cdots, M+1.\nn
\end{align}
\end{subequations}
Problem (P5) is a semidefinite programming (SDP) and hence its optimal solution, denoted by $\{\boldsymbol{\Psi}^*,t^*\}$, can be obtained by using existing solvers such as CVX.  Then the optimal solution to problem (P4) is given by $\boldsymbol{\Phi^*}\!=\!\frac{\boldsymbol{\Psi^*}}{t^*}$.  Since $\boldsymbol{\Phi^*}$, in general, may not be of rank-one, i.e., $\rank(\boldsymbol{\Phi}^*)\!\neq\! 1$, the optimal objective value of problem (P4) is an upper bound of problem (P3) only. In this case, the Gaussian randomization method can be used to obtain a feasible and high-quality suboptimal  solution to problem (P3) based on the higher-rank solution obtained by solving (P4) \cite{wu2019intelligent}, which is denoted by $\boldsymbol{{\hat\theta}}$.

Next, based on the obtained near-optimal passive beamforming  $\boldsymbol{{\hat\theta}}$ with continuous phase shifts, we  construct an initial IRS passive beamforming with discrete phase shifts by using \emph{phase compensation} followed by \emph{phase quantization}. Note that our proposed method below applies to any continuous phase-shift initialization.  Specifically, we observe that a rotated passive beamforming $\boldsymbol{{\bar\theta}}\triangleq e^{-j\angle{{\hat \theta}_1}}\boldsymbol{{\hat\theta}}$ yields the same SINR with $\boldsymbol{{\hat\theta}}$, since $\boldsymbol{{\bar\Phi}}\triangleq\boldsymbol{{\bar\theta}}\boldsymbol{{\bar\theta}}^{H}=\boldsymbol{{\hat\Phi}}\triangleq\boldsymbol{{\hat\theta}}\boldsymbol{{\hat\theta}}^{H}$, which leads to the same objective value of (P4). This useful property allows us to rotate $\boldsymbol{{\hat\theta}}$ to $\boldsymbol{{\bar\theta}}$ such as $\angle{{\bar\theta}}_1=0$ without scarifying the rate performance. Then, for each of the remaining ${\bar\theta}_m, m\in\{2, \cdots, M+1\}$, we directly quantize its phase shift to the nearest discrete value in $\mathcal{F}$, given by
\vspace{-5pt}
\begin{equation}
\omega_{m}=\arg\min_{\omega_m\in\mathcal{F}}\l| e^{j\omega_m}- {\bar\theta}_m\r|.
\end{equation}

Last, we successively refine the passive beamforming based on the initial one with discrete phase shifts. Specifically, in each iteration, we find the optimal discrete phase shift for one sub-surface to maximize the SINR in \eqref{Eq:SINRNew} via one-dimensional search over $\mathcal{F}$, with those of the others being fixed, until the fractional decrease of $\gamma(\boldsymbol{\theta})$ in \eqref{Eq:SINRNew} is less than a sufficiently small threshold. The algorithm is guaranteed to converge since the objective value of (P2) is non-decreasing over the iterations and the optimal objective value of (P2) is upper-bounded by a finite value, i.e.,
\vspace{-5pt}
\begin{align}
\frac{\boldsymbol{\theta}^{H}\boldsymbol{{\tilde H}}\boldsymbol{\theta}}{\boldsymbol{\theta}^{H}\boldsymbol{{R}}_p\boldsymbol{\theta}+1}&=\frac{\boldsymbol{\theta}^{H}\boldsymbol{{\tilde H}}\boldsymbol{\theta}}{\boldsymbol{\theta}^{H}(\boldsymbol{{R}}_p+\frac{1}{M+1}\boldsymbol{I})\boldsymbol{\theta}}\\
&=\boldsymbol{\theta}^{H}\boldsymbol{{X}}\boldsymbol{\theta}\le (M+1)\lambda_{\max} (\boldsymbol{{X}}),
\end{align}
where $\boldsymbol{{X}}\triangleq(\boldsymbol{{R}}_p+\frac{1}{M+1}\boldsymbol{I})^{-1}\boldsymbol{{\tilde H}}$ and $\lambda_{\max}(\boldsymbol{{X}})$ denotes the maximum eigenvalue of $\boldsymbol{{X}}$. Note that compared to the exhaustive search, our proposed successive refinement algorithm greatly reduces the complexity. Specifically, the SDP-based high-quality initialization has a complexity of  $\mathcal{O}((M+1)^{3.5})$ and the successive refinement algorithm given any feasible initialization has a low complexity of $\mathcal{O}(\log(1/{\epsilon})2^b M)$, given the solution accuracy of $\epsilon>0$. Thus, the total complexity of the proposed algorithm is $\mathcal{O}(\log(1/\epsilon)2^b M+(M+1)^{3.5})$, which is practically affordable if $M$ and $b$ are moderate values.
%  . Other initialization methods can also be proposed to further reduce the complexity at the cost of rate performance loss. 

%[complexity]
%\begin{remark}[Orthogonal IRS reflection pattern]\emph{Note that if the IRS reflection pattern is orthogonal, i.e., $\boldsymbol{\Theta}^{H}_p \boldsymbol{\Theta}_p=(M+1)\boldsymbol{I}$, we have $\boldsymbol{\theta}^{H}\boldsymbol{{R}}_p\boldsymbol{\theta}=1$ and hence the optimization problem (P2) reduces to maximizing the channel power gain $\boldsymbol{\theta}^{H}\boldsymbol{{\tilde H}}\boldsymbol{\theta}$. For this simplified problem, by reformulating it into an integer linear programming as in \cite{wu2019beamforming}, the optimal solution can be obtained by using the brand-and-bound method.}
%\end{remark}
\vspace{-5pt}
\section{Numerical Results} 
Numerical results are presented in this section to demonstrate the effectiveness of the proposed training reflection pattern and passive beamforming designs. Under a three-dimensional Cartesian coordinate system in meter (m), we assume that a single-antenna user located at $(20, 50, 0)$  transmits data to a single-antenna AP located at $(20, 0, 0)$. The IRS is equipped with a uniform rectangular array with half-wavelength spacing, which is placed in the $y$-$z$ plane centralized at $(18, 50, 0)$.  For large-scale fading, the path loss exponents of the user-AP, user-IRS and IRS-AP channels are set as $4.5$, $2.2$, and $2.5$, respectively, and the reference channel power gain at a distance of $1$ m is $-30$ dB. All the channels are assumed to experience small-scale Rayleigh fading. 
Other parameters are set as  $P_t=20$ dBm,  $\sigma^2=-79$ dB, $\Gamma=8.2$ dB, and $T_0=40$. 

\begin{figure}[t]
\vspace{4pt}
\begin{center}
\includegraphics[height=4.8cm]{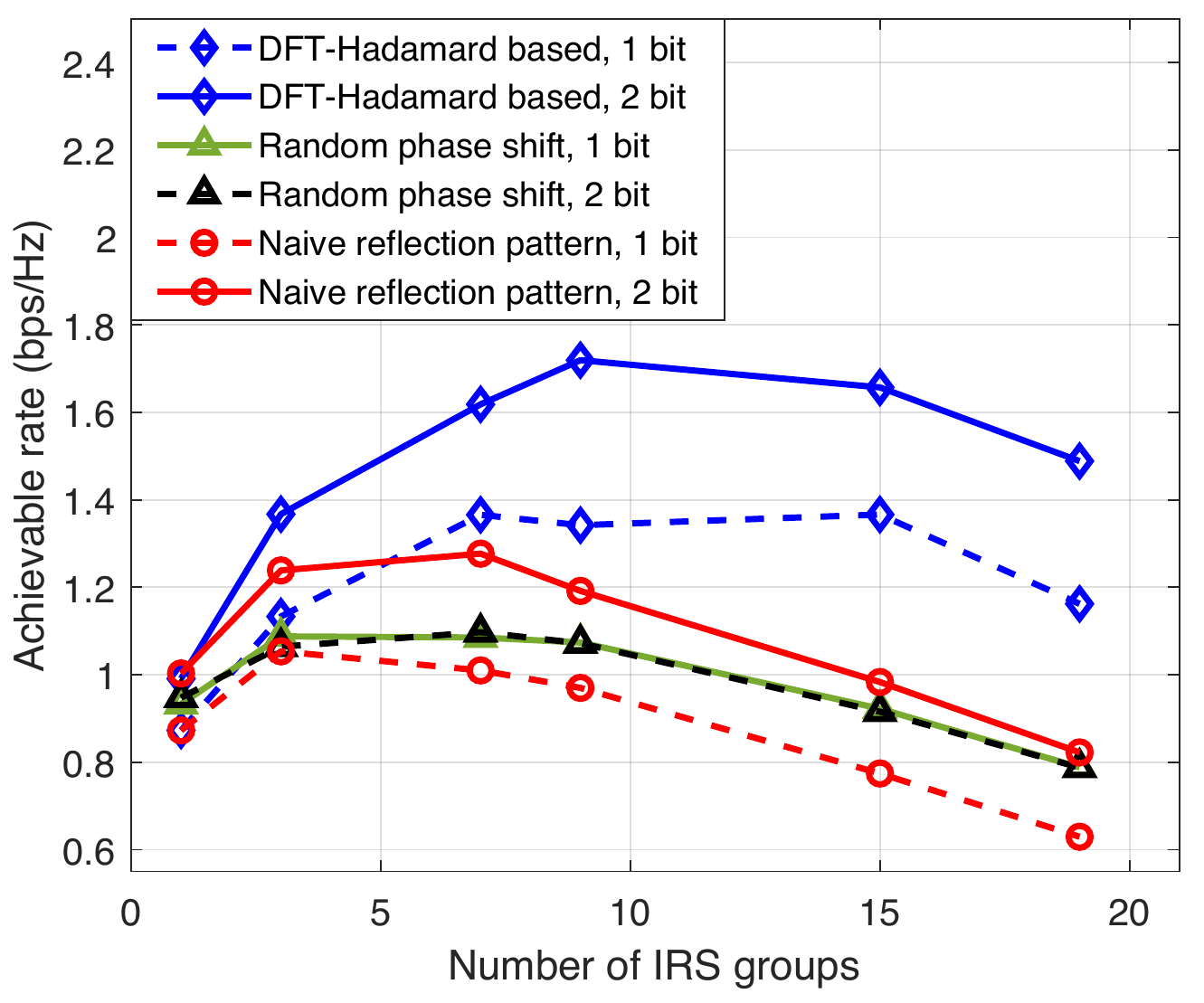}
\caption{Achievable rate versus the number of IRS groups.}
\label{FigGroupRate}
\end{center}
%\vspace{-8pt}
\end{figure}

We first evaluate the effectiveness  of our proposed DFT-Hadamard-based reflection pattern design. The following benchmark schemes are considered: 
1) random phase shift in which the IRS generates $(M+1)$ sets of random reflection coefficients  during the channel-training phase and the AP selects the best set that achieves the largest rate for data transmission; 2) naive reflection pattern whose entries are given in \eqref{Eq:NaiveSch}. Both of the proposed scheme and the naive scheme apply the successive refinement algorithm for designing the passive beamforming at the IRS for data transmission. Fig.~\ref{FigGroupRate} plots the achievable rates of different schemes versus the number of groups with $N=80$. First, it  is observed that there exists a tradeoff between the IRS reflection performance and the number of groups for all schemes, since with too little training the CSI is not accurate enough for the reflection design
to achieve  high passive beamforming gains, while too much training results in less time for data transmission.
Second, our proposed scheme greatly outperforms the benchmark schemes  due to the near-orthogonality of the designed  reflection pattern.  
In addition, one can observe significant rate improvement of the proposed scheme by increasing the resolution of discrete phase shifters  from  $1$-bit  to $2$-bit, whereas the random phase shift scheme shows similar rates  for these two phase-shift levels. Another interesting observation is that the naive reflection pattern achieves lower rates than the random phase shift scheme when $b=1$, but outperforms it when $b=2$. The reason is that, although the training  reflection pattern is the same for the naive scheme with $1$-bit or $2$-bit phase shifters, increasing the resolution of  phase shifters for the passive beamforming  can substantially enhance the rate performance for data transmission.
%due to the reduction of phase quantization error.

\begin{figure}[t]
\vspace{4pt}
\begin{center}
\includegraphics[height=4.8cm]{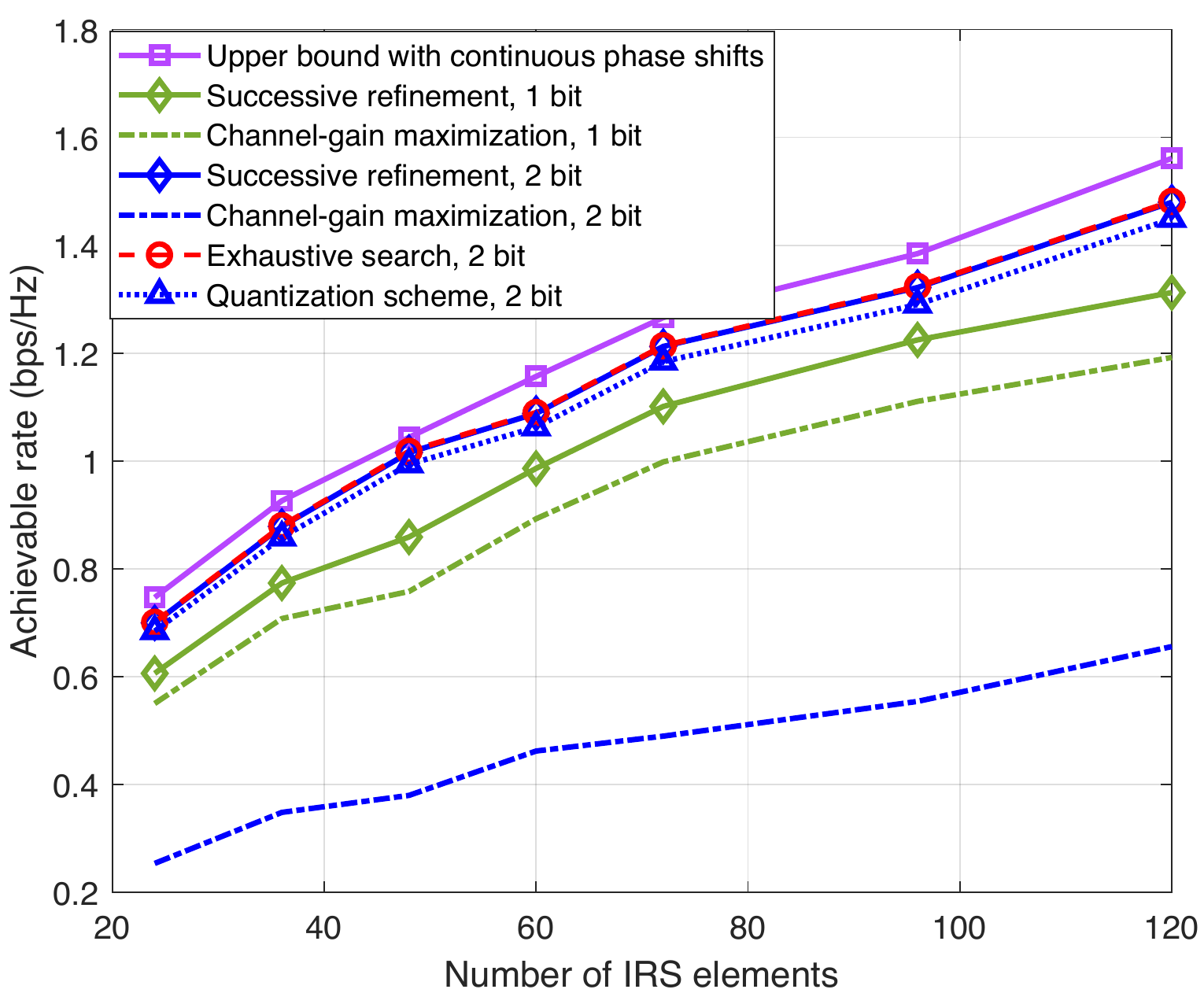}
\caption{Achievable rate versus the number of reflecting elements.}
\label{Fig:RateEle}
\end{center}
\end{figure}

Next, to show the superiority of the proposed successive refinement algorithm for passive beamforming, we consider the following benchmark schemes with fixed $M=4$: 1) upper bound that solves problem (P3) to obtain the near-optimal solution with continuous phase shifts; 2) exhaustive search that searches all possible passive beamforming   with discrete phase shifts for rate maximization; 3) quantization scheme which maps the continuous phase shifts of the upper-bound scheme to their nearest discrete values; 5) channel-gain maximization which neglects the effects of  correlated channel estimation error and only maximizes the channel power gain (i.e. $\boldsymbol{\theta}^{H}\boldsymbol{{\tilde H}}\boldsymbol{\theta}$) instead of the SINR.  All the above schemes apply the proposed DFT-Hadamard-based reflection pattern for channel estimation. In Fig.~\ref{Fig:RateEle}, we compare the achievable rates of different schemes versus the number of reflecting elements. The key observations are made as follows.
First,  the proposed successive refinement algorithm for $b=2$ achieves near-optimal rate performance with the exhaustive search and only suffers from small rate loss as compared to the upper bound. This indicates that an IRS of low-resolution phase shifters  (e.g. $2$-bit) is able to reap most of the passive beamforming gain of the IRS with continuous phase shifts.  Next, the small gap between the successive refinement algorithm and quantization scheme demonstrates the effectiveness of the proposed initialization method for IRS passive beamforming by solving (P5). Moreover, the successive refinement algorithm significantly outperforms the random phase shift scheme, since it designs customized reflection pattern for reducing the channel estimation error and fully exploits CSI for passive beamforming, whereas the random phase shift scheme sacrifices the beamforming gain without combining the signals effectively. Last, it is surprising to observe the substantial rate improvement of the successive refinement algorithm over the channel-gain maximization scheme for  the IRS with $2$-bit phase shifters. The reason can be explained as follows. For the IRS with $4$ sub-surfaces, the proposed reflection pattern is far from orthogonal, making the channel estimation error highly correlated. However, the channel-gain maximization scheme neglects the effects of correlated channel estimation error in the passive beamforming design and thus suffers from considerable rate loss, which is more significant when the IRS has higher-resolution phase shifters due to the suboptimal phase-shift design for data transmission.

\vspace{-5pt}
\section{Conclusion}
In this paper, we considered an IRS-aided single-user communication system with discrete phase shifts, and studied the practical designs of IRS reflection pattern for channel estimation  and passive beamforming for data transmission. For channel estimation, we proposed a new \emph{DFT-Hadamard-based} reflection pattern to minimize the MSE of channel estimation under the constraints of unit-modulus, discrete phase, and full rank. Next, we formulated a rate-maximization problem by designing the IRS passive beamforming based on the reflection-pattern-dependent channel estimation error.  A low-complexity successive refinement algorithm with a properly-designed initialization was proposed to obtain a high-quality suboptimal solution.
 Last, numerical results demonstrated the effectiveness of the proposed reflection pattern and passive beamforming designs.
\vspace{-5pt}
%\bibliographystyle{IEEEtran}
%%\vspace{-4pt}
%\bibliography{BibDesk_File}

\end{document}

%% file: header.tex
\newtheorem{proposition}{\emph{\underline{Proposition}}}

\def\phi{\varphi}

\def\rank{{\operatorname{rank}}}

\def\l{\left}
\def\r{\right}
\def\({\left(}
\def\){\right)}

\def\b0{{\mathbf{0}}}

\newcommand{\tr}{\mathrm{tr}}
\newcommand{\diag}{\mathrm{diag}}

\newcommand{\nn}{\nonumber}